\documentclass[usenatbib]{mn2e}
\usepackage{amsmath}
\usepackage{graphicx}
\usepackage{natbib}
\usepackage{wasysym}
\usepackage{threeparttable}
\usepackage{url}

\newcommand{\cuvir}{CU Vir}
\renewcommand{\tnote}[1]{$^{\textrm #1}$}
\newcommand{\titem}[1]{\item \tnote{#1}}

\footnotesize
\newdimen\digitwidth    
\setbox0=\hbox{\rm0}
\digitwidth=\wd0
\catcode`!=\active
\def!{\kern\digitwidth}
\normalsize
\title{Observations and modelling of pulsed radio emission from CU Virginis}
\author[K.K.Lo et al.]{K.K. Lo$^{1,2,6},$
J.D. Bray$^{3,2},$ 
G. Hobbs$^{2}$, 
T. Murphy$^{1,4,6}$,
B.M. Gaensler$^{1,6}$,
D. Melrose$^{1}$,
\and 
V. Ravi$^{5,2}$,
R.N. Manchester$^{2}$,
M.J. Keith$^{2}$
\\
$^1$ Sydney Institute for Astronomy, School of Physics, The University of Sydney, NSW 2006, Australia\\
$^2$ Australia Telescope National Facility, CSIRO Astronomy and Space Science, P.O. Box 76, Epping, NSW 1710, Australia\\
$^3$ School of Physics, The University of Adelaide, SA 5005, Australia\\
$^4$ School of Information Technologies, The University of Sydney, NSW 2006, Australia\\
$^5$ School of Physics, University of Melbourne, Parkville, VIC 3010, Australia\\
$^6$ ARC Centre of Excellence for All-sky Astrophysics (CAASTRO)
}
\date{}

\pagerange{\pageref{firstpage}--\pageref{lastpage}}
\pubyear{2011}

\begin{document}

\maketitle

\label{firstpage}

\begin{abstract}
We present 13\,cm and 20\,cm radio observations of the magnetic chemically peculiar star CU
Virginis taken with the Australia Telescope Compact
Array. We detect two circularly
polarised radio pulses every rotation period which confirm previous detections.  
In the first pulse, the lower frequency emission arrives before the higher frequency emission and the ordering reverses in the second pulse. In order to
explain the frequency dependence of the time between the two pulses,
we construct a geometric model of the magnetosphere of CU Virginis, and
consider various emission angles relative to the magnetic field
lines. A simple electron cyclotron maser emission model, in which the
emission is perpendicular to the magnetic field lines, is not
consistent with our data. A model in which the emission is refracted through cold plasma in the magnetosphere is shown 
to have the correct pulse arrival time frequency dependence.    

\end{abstract}

\begin{keywords}
radio continuum: stars $-$ stars: individual: CU Virginis, magnetic fields,
rotation $-$ radiation mechanisms: non-thermal  
\end{keywords}

\section{Introduction}

CU Virginis (HD124224, hereafter CU Vir) is one of the best studied
Ap stars. It is very nearby (distance
of only 80\,pc) and is a fast rotator (period of around
0.52 days) \citep{deutsch1952}. Its strong magnetic field (polar magnetic field of $\sim$3\,kG) 
places it in the category of Magnetic Chemically Peculiar (MCP) stars. Like other MCP stars, its effective magnetic
field has been observed to vary with
rotational phase \citep{borra+80}. In addition, MCP stars are known radio
emitters \citep{Linsky1992, Leone1994, dab+87} and their radio emission has been observed
to modulate with rotational phase \citep{Leone1993,Leone1991}. 
This quiescent radio emission is believed
to be gyrosynchrotron radiation emitted by mildly relativistic
(Lorentz factor of $\gamma \leq 2$) electrons trapped in the magnetosphere
\citep{dab+87}.  \citet{Trigilio2004} constructed a three-dimensional model to explain the rotational modulation of the quiescent radio emission from a MCP star and \citet{Leto2006} successfully applied it to CU Vir. 

Circular polarisation in the radio emission was detected at the $10\%$ level by \citet{lut+96} which is expected for the gyrosynchrotron emission. Subsequently, \citet{tll+00} discovered that CU Vir produces two 100\%
circularly polarised radio pulses every rotation period which cannot arise from the gyrosynchrotron emission mechanism. 

The observed radio flux for CU Vir across one rotation period is hereafter described as the
``pulse profile''.  The two components in the pulse profile are
referred to as the ``leading'' and ``trailing'' pulses.
Similar pulse profiles have been observed from CU Vir in the 13\,cm and 20\,cm
observing bands \citep{tll+00, tlu+08,trigilio2011,rhw+10} where observations were separated by
almost a decade. CU Vir is the only MCP star that has been observed to emit radio pulses. 

\citet{tll+00} proposed electron cyclotron maser (ECM) as
the emission mechanism for the pulsed radio emission. Plasma radiation due to Langmuir waves was ruled out  
\citep{tlu+08} because of the low electron density in the emission region as modelled by \citet{Leto2006} and the observation that requires the radiation to be tightly beamed. 
ECM can produce radiation with high directivity and 100\%
circular polarization. In the ``loss cone" model of ECM \citep{melrose1982}, ECM
emission forms a hollow cone at a large angle ($70^{\circ}$ to $85^{\circ}$) to
the magnetic field lines, and occupies a narrow frequency band close to a
harmonic of the cyclotron frequency. Although ECM is narrow band, the relatively wide 
band of the observation can be explained if the emission region spans a range of 
magnetic field strengths. 

A simple ECM model does not explain all the features of the pulsed radio
emission from CU Vir. For instance, in the 20\,cm observing band, the leading and
trailing pulses are separated by $\sim5$\,hours. In the 13\,cm
observing band, the leading pulse was not seen in observations by
\citet{tlu+08}, but was seen a decade later by \citet{rhw+10}. In
contrast to the variability of the leading pulse, the trailing pulse
shows remarkable stability over more than a decade in both observing bands. 
The trailing pulse at 13\,cm is always observed to arrive
earlier than the corresponding 20\,cm pulse.  \citet{tlu+08}
postulated that this has a geometric explanation, which is related
to where the pulse originates in the magnetosphere. The discovery from
\citet{rhw+10} that the leading 13\,cm pulse arrives later than the
corresponding 20\,cm pulse, agrees with this interpretation and
suggests that both pulses come from the same magnetic
pole of the star. However, until now, existing data do not provide enough frequency coverage to
study the pulse profile in detail as a function of observing
frequency. Here we present new data with more complete frequency coverage over
the 13\,cm and 20\,cm observing bands and demonstrate with our   
modelling that an ECM emission model, without propagation effects, cannot
explain the observations. In a recent paper, \citet{trigilio2011} suggest  
the frequency dependence of the pulse arrival time arises from frequency
dependent refraction of the
radiation through the stellar magnetosphere. In this paper, we have analysed the frequency 
dependent refraction model in more detail.      

The structure of the paper is as follows. In \S2 we present the
observing and analysis techniques used to detect time varying emission
from CU Vir with the Australia Telescope Compact Array (ATCA) and
present our observational results. In \S3 we discuss our modelling
and show how it relates to our observations.
 
\section{Observations and Data Analysis}
\label{sec:obs}
We observed CU Vir with the ATCA on six separate occasions (2009 December 23 to 26, 2010 May 19, 2010 June 15). The
observations in December 2009 were only two hours each in length because
they were taken during unallocated telescope time. For these short
observations, we predicted the arrival time of the trailing pulse
using the ephemeris from \citet{rhw+10} and timed our observation
accordingly. We were allocated time for the May and June 2010 epochs
and hence were able to observe for about nine hours continuously. A
summary of the observation parameters is given in Table
\ref{tab:atcaParams}. 

All our observations were taken with the Compact Array Broadband
Backend (CABB; Wilson 2011\nocite{wilson2011}) which offers a
bandwidth of up to 2\,GHz per intermediate frequency 
and polarisation. However, the maximum bandwidth in
our 20 and 13\,cm observing bands was limited to around
700\,MHz because our observations were taken before the upgrade
was fully completed. Furthermore, it was not possible to observe at 20 and
13\,cm simultaneously. To maximise the apparent bandwidth, we switched
between the 20 and 13\,cm bands every few minutes in the 2009 Dec 25 to 26 and
2010 May 19 epochs. Complex visibility data in four polarisations were recorded for 15 
baselines with a sampling time of 10\,s.

We also made use of archival (pre-CABB) ATCA data to supplement our
analysis. Altogether, we analysed eight epochs of data at 20\,cm and
seven epochs at 13\,cm as summarised in Table \ref{tab:atcaParams}.
In column order we provide the date of observation, the observing
frequencies used\footnote{We refer to the 1384\,MHz pre-CABB and 1503\,MHz CABB
frequencies as the 20\,cm band; and the pre-CABB 2496\,MHz and CABB 2335\,MHz
frequencies as the 13\,cm band}, the available bandwidth, the configuration of the
ATCA\footnote{In all cases a maximum baseline of 6\,km
was used. The various labels (6A, 6C and 6D) refer to different
antenna spacings, which are defined at
\url{http://www.narrabri.atnf.csiro.au/operations/array_configurations/configurations.html}},
the total integration time, the phase calibrator used and
a reference for the data. 

\begin{table*}
\caption{Summary of ATCA observation parameters} 
\label{tab:atcaParams}
\begin{tabular}{l l l l l l l l}
\hline
Date & Start time & Frequency & Bandwidth & ATCA configuration & Integration time&Phase calibrator& Reference\\
& UT & (MHz) & (MHz) & & (hours) & &\\
\hline
1999 May 29 & 06:57 & 1384, 2496 & 128, 128 & 6A & 9.3 & B1406$-$076&\citet{tlu+08}\\
1999 Aug 29 & 01:12 & 1384, 2496 & 128, 128 & 6D & 9.0& B1406$-$076&\citet{tlu+08}\\
2008 Oct 30 & 21:06 & 1384, 2368 & 128, 128 & 6A & 9.0 & B1416$+$067&\citet{rhw+10}\\
2009 Dec 23 & 18:13& 1503 & 742 & 6A & 1.2 & B1406$-$076& This work\\
2009 Dec 24 & 18:11& 2335 & 678 & 6A & 1.6 & B1406$-$076& This work\\
2009 Dec 25 & 20:01 & 1503, 2335 & 742, 678 & 6A & 1.9 & B1406$-$076 & This work\\
2009 Dec 26 & 20:13 &1503, 2335 & 742, 678 & 6A & 2.4 & B1406$-$076 & This work\\
2010 May 19 & 08:02& 1503, 2335 & 742, 678 & 6C & 8.2 & B1416$+$067& This work\\
2010 Jun 15 & 06:41&1503 & 742 & 6C & 9.2 & B1402$-$012 & This work\\
\hline
\end{tabular}


\end{table*} 

We calibrated our data using the standard \textsc{miriad} software package
\citep{sault+95}\footnote{\url{http://www.atnf.csiro.au/computing/software/miriad/}}. The
flux amplitude scale and the bandpass response were determined from
the backup ATCA primary calibrator 0823$-$500. Observations of a bright
compact radio source (as listed in Table \ref{tab:atcaParams}) for
1.5 minutes every 20 minutes were used to calibrate the complex gains
and leakage between the orthogonal linear feeds in each antenna. Radio
frequency interference was removed from the visibilities using the
flagging tool \textsc{mirflag}.

To obtain light curves, we first imaged the field in the standard
way. Then, we took the \textsc{clean} components of each source, masking the location of CU Vir, and subtracted them from the visibility data using the \textsc{miriad} task \textsc{uvmodel}. We then
shifted the phase centre of the visibilities to the location of CU Vir and 
measured the flux density directly from the visibilities by vector averaging their real 
components in time bins of three minutes.  

\subsection{Pulse profile} 

\begin{figure}
   \centering
   \includegraphics[width=90mm]{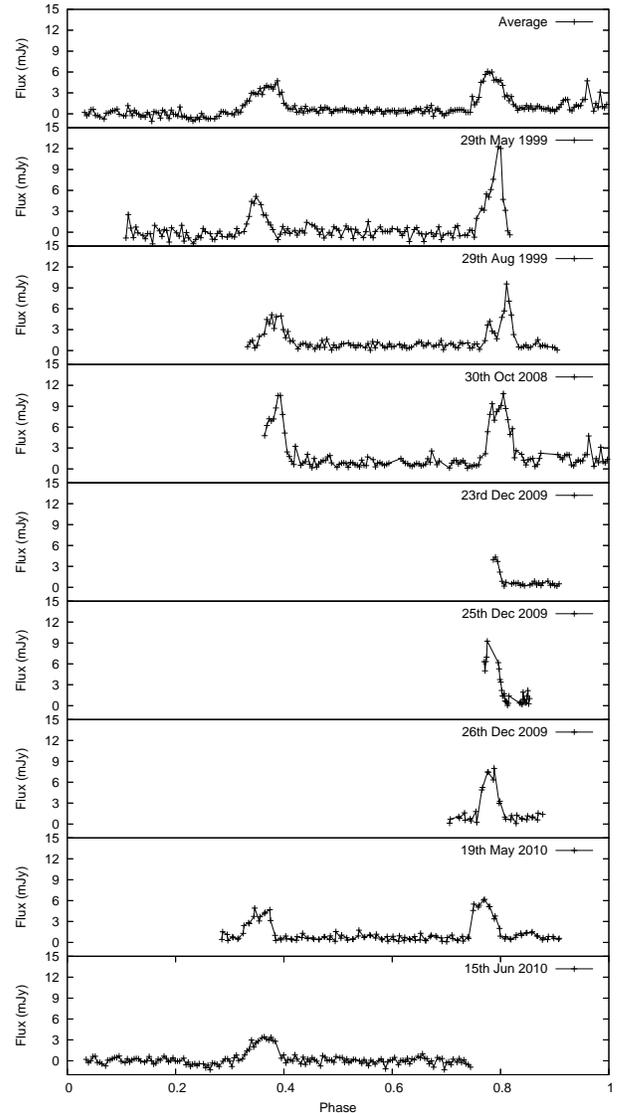}
   \caption{20\,cm Stokes V pulse profiles from all epochs listed in Table \ref{tab:atcaParams} aligned using the rotation period from \citet{trigilio2011}.}   
   \label{fig:20cm_V}
\end{figure}

\begin{figure}
   \centering
   \includegraphics[width=90mm]{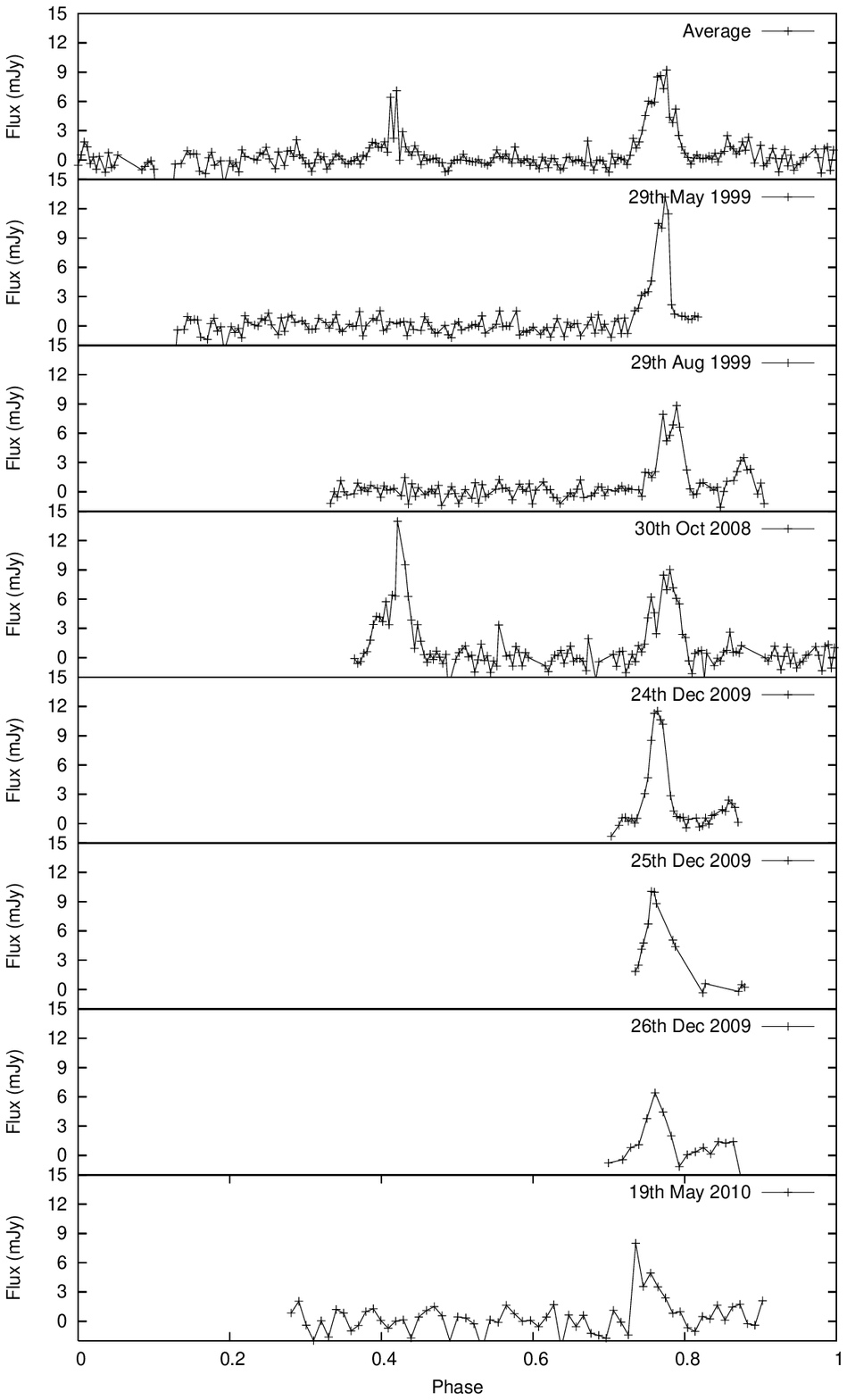}
   \caption{13\,cm Stokes V pulse profiles from all epochs listed in Table \ref{tab:atcaParams} aligned using the rotation period from \citet{trigilio2011}}
   \label{fig:13cm_V}
\end{figure}

The 20 and 13\,cm Stokes V pulse profiles from our observations, along
with other archival epochs, are shown in Figures~\ref{fig:20cm_V} and
\ref{fig:13cm_V} respectively. The Stokes V flux densities are plotted against
rotational phase using the ephemeris from \citet{trigilio2011} and adjusted for
the rotation of the Earth around the Sun. We have used the rotation period, 
$P = 0.52071601$\,days from \citet{trigilio2011} to align the light curves and set phase
zero, in HJD, to be $2450966.3601 + \textrm{EP}$, where E is the number of epochs. As can be seen in Figures
\ref{fig:20cm_V} and \ref{fig:13cm_V}, the rotation period is a good fit to our
data as the pulses are aligned in phase. 

\begin{figure}
   \centering
   \includegraphics[width=60mm, angle=270]{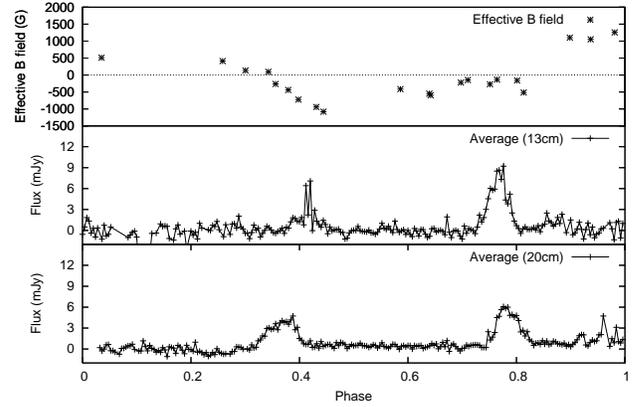}
   \caption{Top: Effective magnetic field B$_\textrm{eff}$ from \citet{prm+98} vs. rotation phase. A negative B$_\textrm{eff}$ means the field lines are directed into the star. Middle and bottom: Average Stokes V pulse profiles at 13\,cm and 20\,cm. If we assume CU Vir has a dipolar field structure, then the magnetic South pole is closest to our line of sight during the shorter of the two pulse separations.}
   \label{fig:bfields}
\end{figure}

In the May 2010 epoch, we detected 
both the leading and trailing pulses. The time between them is
5.20$\pm$0.05 hours in the 20\,cm observing band. This is similar to the pulse
separation time determined by \citet{tll+00} and \citet{rhw+10} in
earlier epochs. 
In Figure~\ref{fig:bfields}, we have
juxtaposed measurement of the effective magnetic field from \citet{prm+98} with the average Stokes V pulse profiles. 
Assuming a dipolar field structure, and take negative effective magnetic field to mean the field lines point into the star, the magnetic South pole is closest to our line of sight during the shorter of the two pulse separations.  

We did not detect the 13\,cm leading pulse, which
had been detected in 2008. Although the trailing pulse was detected in all
epochs for which we have data at the relevant phase, 
the peak flux density varied between 6-12\,mJy in the 20\,cm
observing band, and 7-13\,mJy in the 13\,cm observing band. With
our data we can constrain the frequency at which the pulse disappears 
to be between 1.9 and 2.3\,GHz in the 19 May 2010 epoch.

\begin{figure}
   \centering
   \includegraphics[width=90mm]{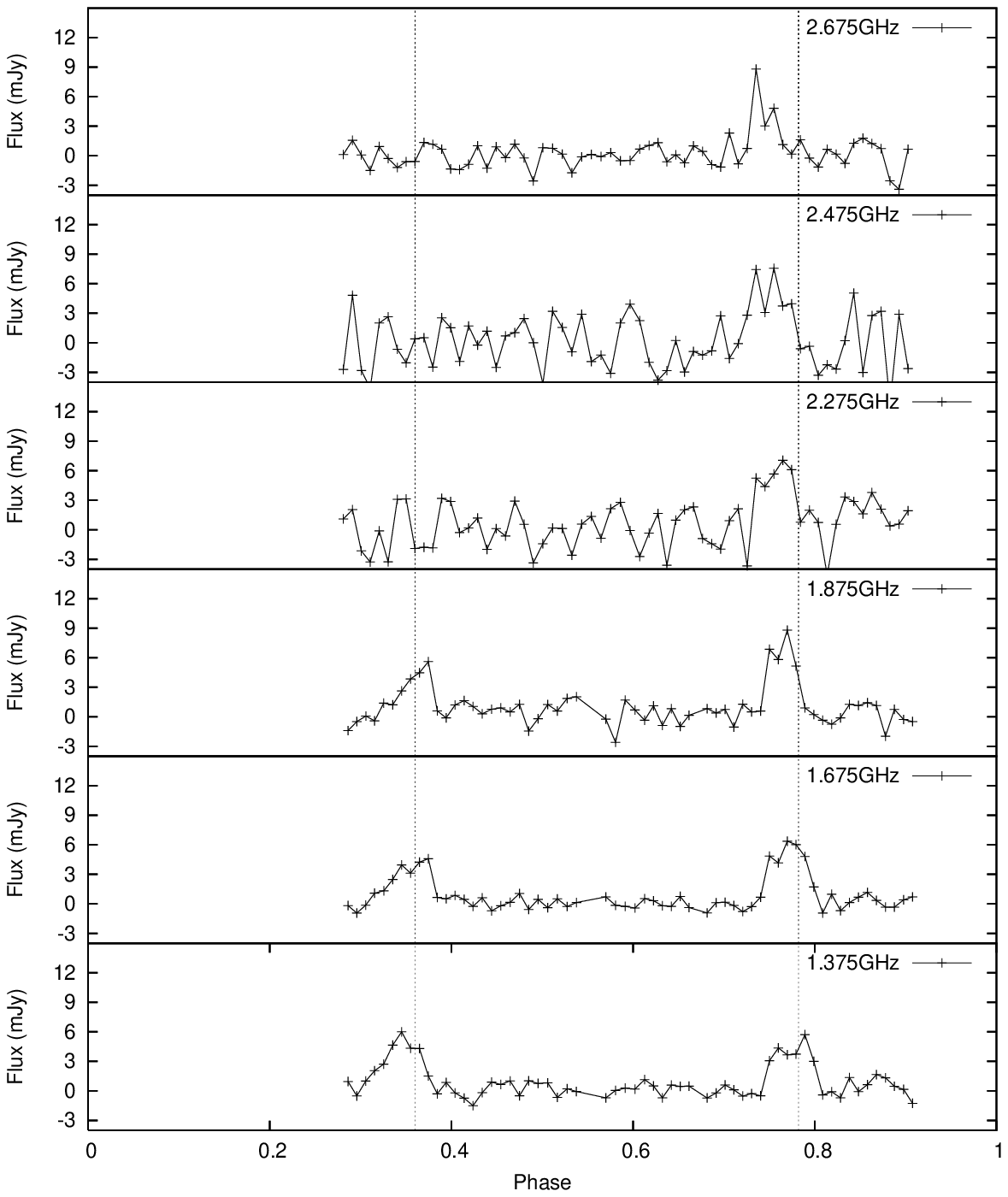}
   \caption{Stokes V pulse profile from the 19 May 2010 epoch. Each panel
represents the average of 200\,MHz bandwidth. The
noise is higher for the three pulse profiles in the 13\,cm observing band compared to 20\,cm observing band because more channels were removed due to RFI.}
   \label{fig:wideband_V}
\end{figure}

\begin{figure}
   \centering
   \includegraphics[width=90mm]{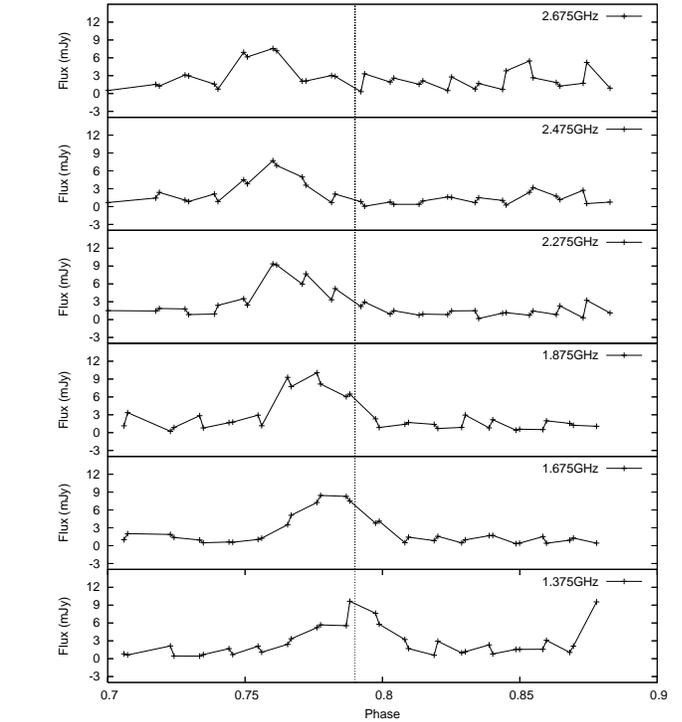}
   \caption{Stokes V pulse profile from the 26 Dec 2009 epoch. It is clear that the 1.4\,GHz pulse lags the 2.7\,GHz pulse by $\sim0.03$
phase.}
   \label{fig:wideband_V_2009}
\end{figure}

Figure~\ref{fig:wideband_V} shows the pulse profiles, in 200\,MHz
channels, for the 19 May 2010 epoch. 
The pulse arrival phase is dependent upon the observing frequency. This dependence is clearly observed for the trailing pulse in the 26 Dec 2009 epoch (Figure~\ref{fig:wideband_V_2009}). This frequency dependence
is stable over at least six months since we observe similar phenomenon
in the trailing pulse in the 26 Dec 2009 epoch.
We determined the arrival phase
of the pulses by cross correlating the single epoch pulse and the average pulse,
and finding the phase of the peak of the cross correlation. 
The error of the
pulse arrival phase is the variation of the parameter required to increase the
reduced $\chi^2$ between the single epoch pulse and the average pulse by one
unit. Table~\ref{tab:toa} summarises the arrival phase for the 19 May 2010 epoch.

\begin{table}
\caption{Phases of the leading and trailing pulse in the 19 May 2010 epoch} 
\label{tab:toa}
\begin{tabular}{l l l l}
\hline
Frequency &Leading pulse&Trailing pulse&Pulse separation\\
(MHz) & arrival phase & arrival phase & (hours) \\
\hline
1.375 & $0.360\pm0.002$ & $0.782\pm0.002$ & $5.27\pm0.05$ \\
1.675 & $0.366\pm0.002$ & $0.778\pm0.002$ & $5.15\pm0.05$ \\
1.875 & $0.372\pm0.006$ & $0.776\pm0.006$ & $5.1\pm0.1$ \\
2.275 & - & $0.76\pm0.02$ & - \\
2.475 & - & $0.76\pm0.02$ & - \\
2.675 & - & $0.756\pm0.006$ & - \\
\hline
\end{tabular}


\end{table}

\section{Modelling}

Models for the origin of the pulsed emission from \cuvir\ have been proposed by \citet{tll+00}, \citet{tlu+08} and \citet{trigilio2011}.  We perform numerical simulation of each of these models to produce a pulse profile, and compare these with our observations.

\subsection{The magnetosphere of \cuvir}

\begin{figure*}
 \includegraphics[width=0.8\linewidth]{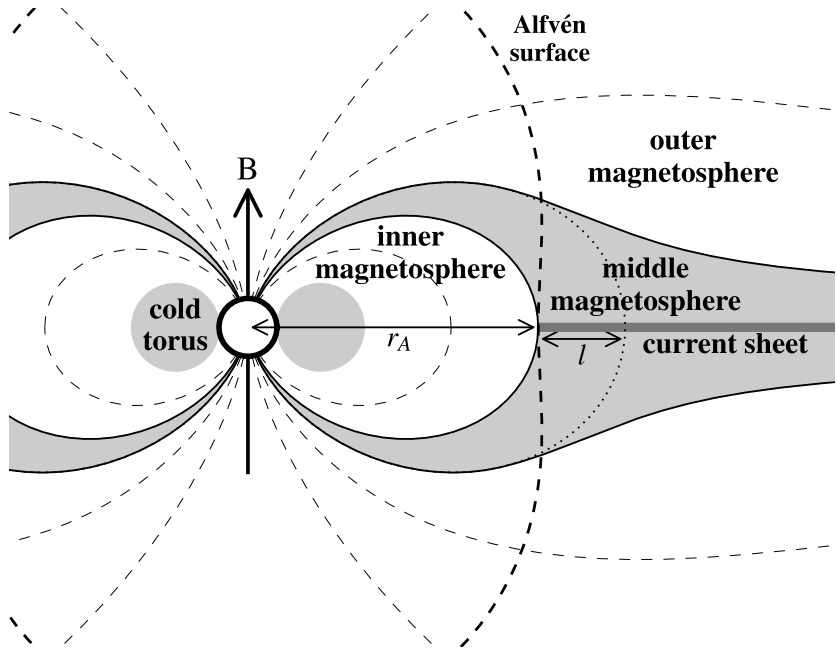}
 \caption{Cross-section of the magnetosphere of \cuvir\ used in our modelling (not to scale).
The magnetic field is assumed to be dipolar within the Alfv\'en surface (heavy dashes), with equatorial radius
$r_A$. Beyond this radius, field lines (light dashes) are distorted by the stellar
wind. Electrons are accelerated in equatorial current sheets beyond the Alfv\'en
radius, and travel along field lines within the middle magnetosphere (shaded),
which divides the inner and outer magnetospheres.  The parameter $l$ describes
the extent of the middle magnetosphere, indicating the position of the
projection (dotted) of the field line marking its outer boundary within the
Alfv\'en surface.  Cooled gas from the stellar wind accumulates in a torus (shaded) near the star.}
 \label{fig:model}
\end{figure*}

The emission models described here are based on a stellar magnetosphere model proposed by \citet{andre1988}, shown in Figure~\ref{fig:model}.  The cold torus, containing trapped material from the stellar wind, is an extension by \citet{Trigilio2004}.

In this model, electrons are accelerated by magnetic reconnection in the magnetic equatorial plane of the star, and propagate along field lines towards the magnetic poles.  They are reflected through magnetic mirroring; however, electrons with sufficiently small pitch angles will instead intersect the surface of the star.  This results in a loss-cone anisotropy in the pitch angles of this population of electrons, allowing them to produce radio emission through the electron cyclotron maser mechanism~\citep{tll+00}.

ECM emission occurs close to a harmonic of the cyclotron frequency: $\nu = s \nu_{cyc}$, where $s$ is the harmonic number.  We assume the emission to be in the second harmonic, $s = 2$, because the fundamental, $s = 1$, is generally suppressed by gyromagnetic absorption of the thermal plasma in weaker magnetic fields as the radiation escapes, and the growth rates for the higher harmonics are too low for the intensity to be significant \citep{melrose1982}.

The sense of the polarised emission (positive Stokes V; right circular polarisation) indicates that it originates entirely from the north magnetic pole of the star.  This may be explained by the presence of a quadrupole component in the magnetic field of \cuvir~\citep{tll+00}.  Consequently, we consider only the north magnetic hemisphere in our modelling.

\subsection{Model parameters}

Table~\ref{tab:params} lists the parameters we use for our modelling.  The obliquity of the
magnetic axis was calculated by \citet{tll+00} from measurements of the variation in effective magnetic field strength given an inclination $i$.  \citet{Leto2006} derived the Alfv\'en radius and thickness of the magnetosphere by fitting multifrequency radio flux densities to a three-dimensional magnetospheric model simulating the quiescent gyrosynchrotron emission. The dipole moment is calculated from a value of 3\,kG for the magnetic field strength at both poles from
\citet{tll+00}.  \citet{hatzes1997} suggest an off-centre dipole model, with different surface magnetic field strengths at each pole, but these values give a similar dipole moment\footnote{Since the star is treated here simply as a rotating dipolar field, the actual position of its surface is not required for the modelling.}. We use only the central value for each parameter, ignoring the uncertainty.

\begin{table}
 \centering
 
  \caption{Adopted parameters for \cuvir.}
  \label{tab:params}
  \begin{tabular}{p{4cm}cc}
  \hline 
  Parameter & Symbol & Value \\ 
  \hline
   Rotation axis inclination [degrees] & $i$ & $43\pm7$\,\tnote{a} \\
   Magnetic axis obliquity [degrees] & $\beta$ & $74\pm3$\,\tnote{a} \\
   Stellar radius [$R_{\astrosun}$] & $R_*$ & $2.2$\,\tnote{b} \\
   Equatorial Alfv\'en radius [$R_*$] & $r_A$ & $15\pm3$\,\tnote{c} \\
   Thickness of middle magnetosphere [$R_*$] & $l$ & $1.5$\,\tnote{c} \\
   Dipole moment [A\,m$^2$] & $m$ & $5.4 \times 10^{33}$\,\tnote{d} \\
   \hline
  \end{tabular}
  \begin{tablenotes}
   \titem{a} \citet{tll+00} 
   \titem{b} \citet{north1998}
   \titem{c} \citet{Leto2006}
   \titem{d} see text
  \end{tablenotes}
\end{table}

\subsection{Emission from field lines parallel to the magnetic axis}
\label{sec:paramodel}

\begin{figure*}
 \centering
 \begin{minipage}{0.45\linewidth}
  \centering
  \includegraphics[width=\linewidth]{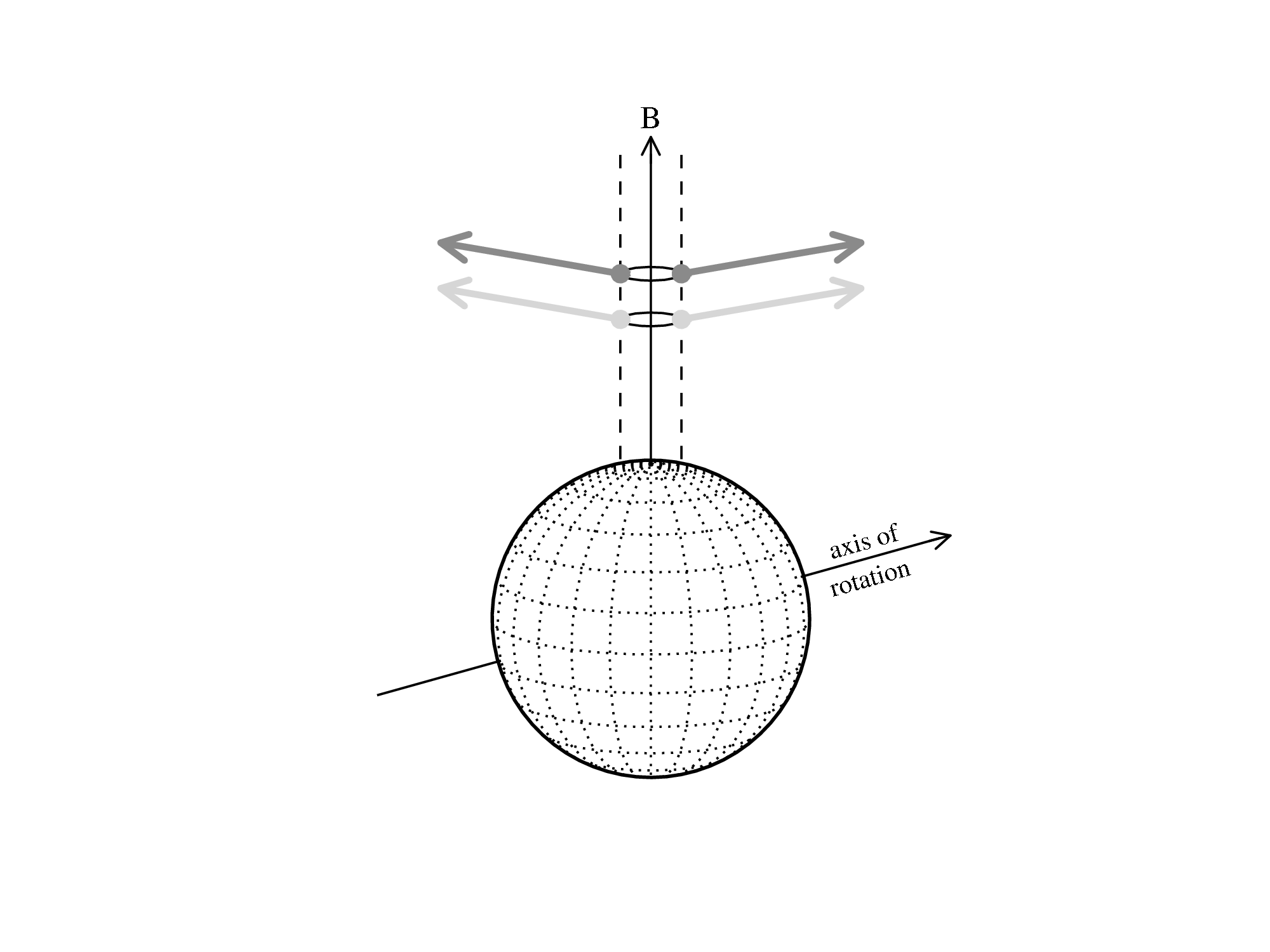}
 \end{minipage}
 \begin{minipage}{0.45\linewidth}
  \centering
  \includegraphics[width=\linewidth]{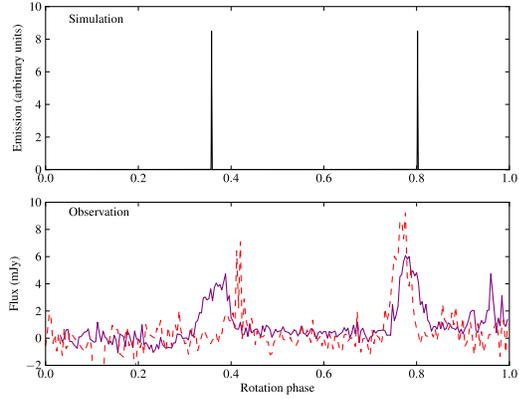}
 \end{minipage}
 \caption{Schematic diagram (left), and pulse profile of emission (right) for
the model of ECM emission from field lines (dashed) parallel to the magnetic
axis (section \ref{sec:paramodel}).  The simulated emission in the 20\,cm (dark arrows) and 13\,cm (light
arrows) bands arises from different locations, but is emitted in the same directions, hence
producing identical pulse arrival times (top right), in contrast to the
averaged observed pulse profile (bottom right). Dotted and solid lines represent the 13\,cm and 20\,cm bands respectively.}
 \label{fig:modelA}
\end{figure*}

The first model for the pulsed emission from \cuvir\ is from \citet{tll+00}.  The emission is assumed to be directed as a hollow cone, making an angle of 85$^\circ$ with the local magnetic field.  At this point, it was believed that the population of high-energy electrons would approach the star along field lines almost parallel to the magnetic axis, as shown in Figure \ref{fig:modelA} (left panel).

Our simulation procedure is as follows:

\begin{enumerate}
 \item \label{it:para_start} Select a rotation phase $\phi$.

 \item Determine the magnetic latitude $\theta_E$ of emission in the direction of the Earth, using:
  \begin{equation}
   \sin\theta_E = \sin \beta \sin i \cos \phi + \cos \beta \cos i
   \label{eqn:rotation}
  \end{equation}
 See appendix~\ref{sec:rot_appendix} for details.

 \item \label{it:para_getem} Determine the strength of emission in this direction.  The emission pattern is a hollow cone making an angle of $85^\circ$ with the north magnetic axis; i.e., at a magnetic latitude of $+5^\circ$.  We assume the emission to have a narrow Gaussian profile centred on this latitude.

 \item Repeat steps \ref{it:para_start}-\ref{it:para_getem} for different rotation phases to form a pulse profile.
\end{enumerate}

A phase of $\phi = 0$ under the definition used in equation \ref{eqn:rotation} corresponds to the point in the rotation of CU Vir when its north magnetic pole is most closely directed towards us.  Therefore, we shift the phase by 0.08 to match the ephemeris as described in section \ref{sec:obs}.   We do not attempt to predict the absolute flux density, but instead scale the peak of the simulated pulse profile to match our observations.

The result of this simulation is shown in Figure \ref{fig:modelA} (right panel).  This model correctly predicts the arrival times of the pulses.  However, it does not fit the width of the pulses, nor their frequency dependence (which had not been observed when this model was developed).

\subsection{Emission from dipolar magnetic field lines}
\label{sec:dipolemodel}

\begin{figure*}
 \centering
 \begin{minipage}{0.45\linewidth}
  \centering
  \includegraphics[width=\linewidth]{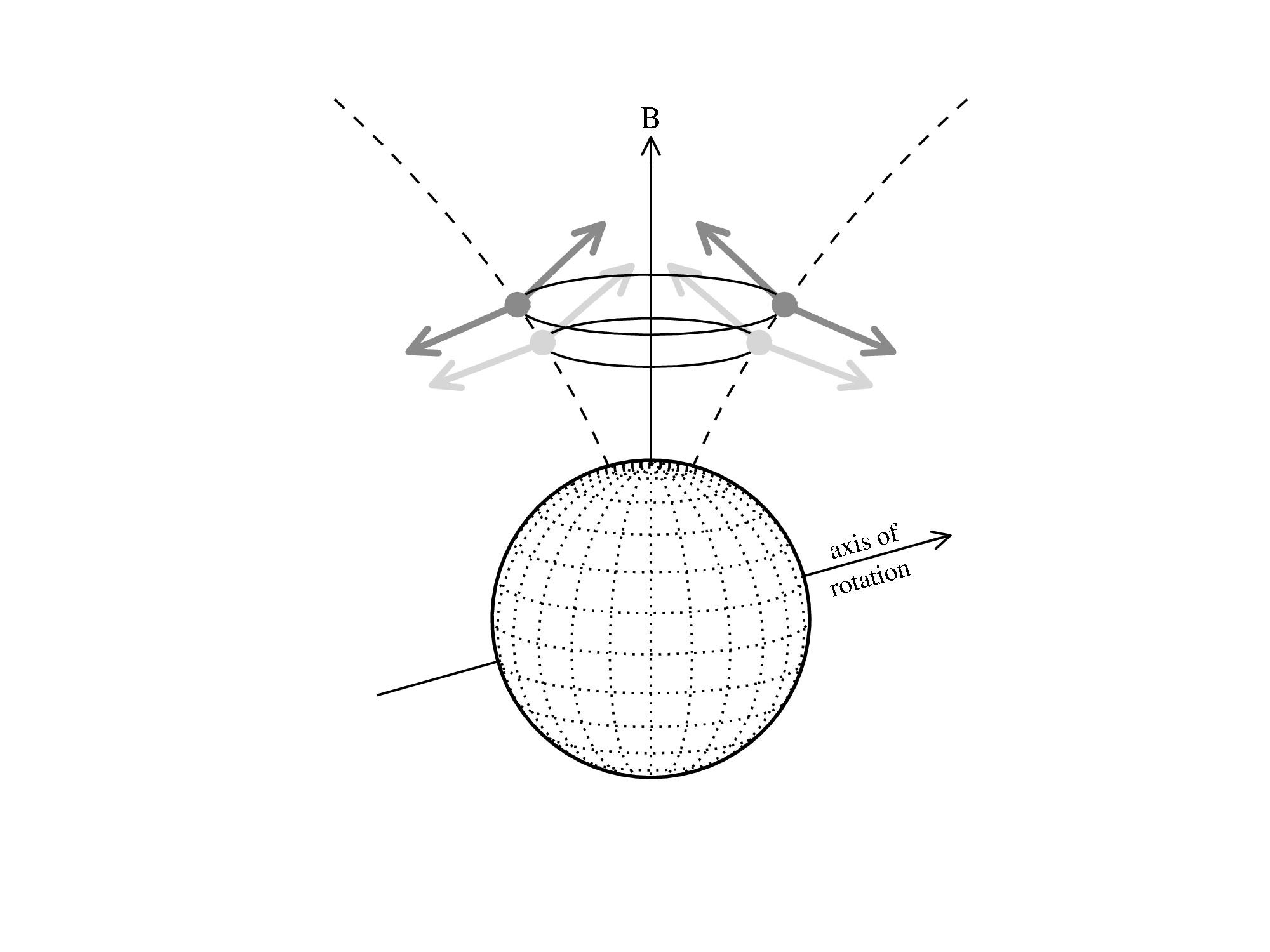}
 \end{minipage}
 \begin{minipage}{0.45\linewidth}
  \centering
  \includegraphics[width=\linewidth]{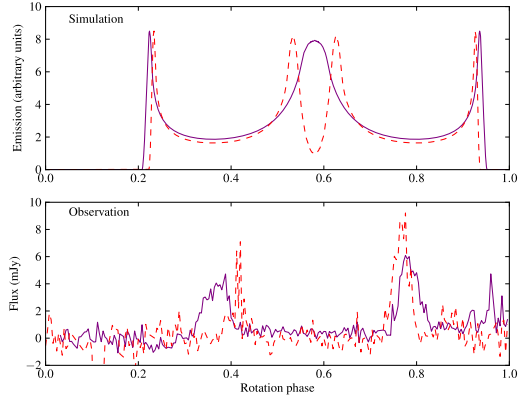}
 \end{minipage}
 \caption{Schematic diagram (left), and pulse profile of emission (right) for
the model of ECM emission from field lines diverging from the magnetic axis (section \ref{sec:dipolemodel}.
The regions in the magnetosphere where emission occurs are marked with solid
rings, and the dark and light arrows correspond to emission in the 20\,cm and
13\,cm bands. The simulated emission (top right) is spread over a very broad range of magnetic
latitudes, resulting in an extended pulse profile quite distinct from our
averaged observations (bottom right). Dotted and solid lines represent the 13\,cm and 20\,cm bands respectively.}
 \label{fig:modelB}
\end{figure*}

This model for the pulsed emission from \cuvir\ is from \citet{tlu+08}.  It incorporates results from \citet{Leto2006}, who determined parameters of the magnetosphere, including the Alfv\'en radius $r_A$.  These values imply that the field lines of the middle magnetosphere near the star are not parallel to the magnetic axis.  We assume them to have a dipolar configuration, as shown in Figures \ref{fig:model} and \ref{fig:modelB} (left panel).

\begin{figure}
  \includegraphics[width=0.95\linewidth]{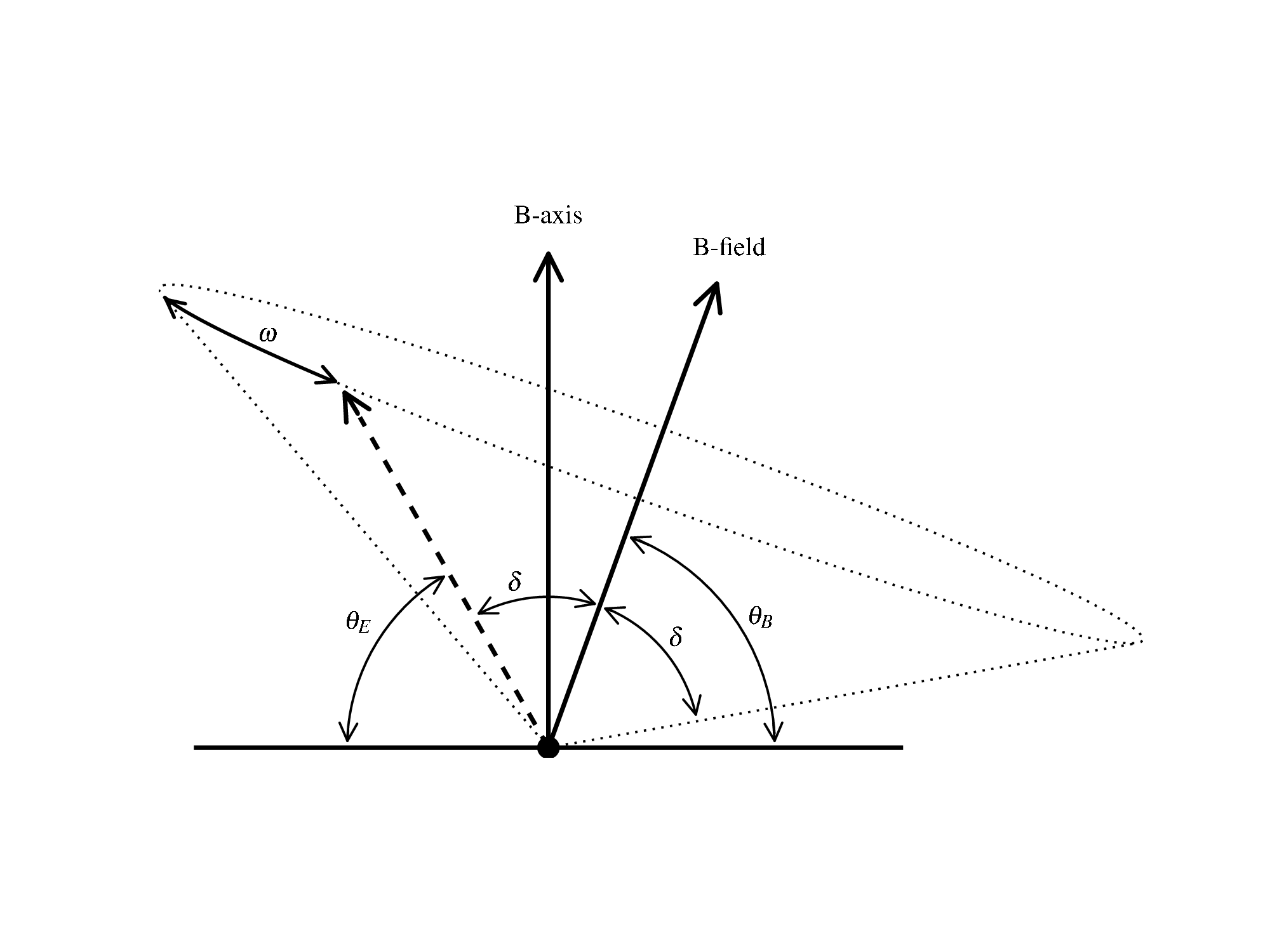}
  \caption{Emission in a hollow cone (dotted) centred on a magnetic field vector
(`B-field') at a magnetic latitude of $\theta_B$.  The half-opening-angle of the
cone is $\delta$.  A sample emission vector is shown as a dashed arrow.  This
vector may be parameterised with the angle $\omega$, which describes its
position on the cone.  Equation \ref{eqn:cone} describes the relationship
between $\omega$ and the magnetic latitude of the emission, $\theta_E$.}
  \label{fig:cone_diagram}
\end{figure}

Our simulation procedure is as follows:

\begin{enumerate}
 \item \label{it:dipole_start} Select a rotation phase and convert to magnetic latitude, as in section \ref{sec:paramodel}.

 \item Select a point on the magnetic equator at radius $r_A$.

 \item From this point, trace along the field line towards the north magnetic pole, following the path that would be taken by an electron in the middle magnetosphere.

 \item At each point along the field line, determine the cyclotron frequency.  When the cyclotron frequency is equal to half of the required emission frequency (corresponding to emission at the second harmonic), stop and note the magnetic latitude $\theta_B$ of the direction of the magnetic field at this point.

 \item \label{it:dipole_decone} The emission is assumed to be directed as a hollow cone centred on an axis with magnetic latitude $\theta_B$.  We determine the parameter $\omega$ which describes the position of the emission vector on this cone, as shown in Figure \ref{fig:cone_diagram}:

\begin{equation}
\cos \omega =
 \frac{\sin \theta_E - \cos \delta \sin \theta_B}
      {\sin \delta \cos \theta_B}
\label{eqn:cone}
\end{equation}

 \item \label{it:dipole_getem} If there was no solution for $\omega$, the emission power $P$ in this direction is zero.  Otherwise, the emission power per solid angle $\Omega$ is:
  \begin{equation}
   \frac{dP}{d\Omega} \propto \frac{1}{ \sin \omega \cos \theta_B \sin \delta }
   \label{eqn:emission}
  \end{equation}
  This formula is obtained by assuming the emission power to be equally distributed around the cone.  See appendix~\ref{sec:cone_appendix} for details.

 \item \label{it:dipole_phase} Repeat steps \ref{it:dipole_start}-\ref{it:dipole_getem} for different rotation phases to form a pulse profile.

 \item \label{it:dipole_cone} Repeat steps \ref{it:dipole_start}-\ref{it:dipole_phase} for different values of $\delta$.  Sum the resulting profiles, weighting them with a narrow Gaussian function centred on $\delta = 85^\circ$.

 \item \label{it:dipole_freq} Repeat steps \ref{it:dipole_start}-\ref{it:dipole_cone} for emission frequencies across the observation band.  Sum the resulting profiles, weighting them equally.

 \item \label{it:dipole_rad} Repeat steps \ref{it:dipole_start}-\ref{it:dipole_freq} for starting points on the magnetic equator with radii between $r_A$ and $r_A + l$.  Sum the resulting profiles, weighting them equally.
\end{enumerate}

The phase and height of the pulse profile are then adjusted as in section \ref{sec:paramodel}.  The Gaussian function used in step \ref{it:dipole_cone} is very narrow (width $< 0.2^\circ$), intended only to smooth over the step size in other parameters, so the width of the resulting profile is due to other aspects of the model.

Results are shown in Figure \ref{fig:modelB} (right panel).  The simulated pulses are too wide to fit the observations, and display a distinct double-peaked structure which is absent in the real data.  There is some frequency dependence, but it affects the pulse width (narrower at higher frequency) rather than the arrival time.

\subsection{Refracted equatorial emission}
\label{sec:tangentmodel}

This model for the pulsed emission from \cuvir\ is from \citet{trigilio2011}.  Rather than a hollow cone centred on the local field line, the ECM emission is assumed to be directed perpendicular to the local field line, and only in the two directions which are also parallel to the magnetic equatorial plane.  For each longitude in the magnetic equatorial plane, this results in emission being visible from two points in the emission region.

The key point of this model is that it includes the effect of refraction as the radiation crosses the boundary of the cold torus region (Figure \ref{fig:model}).  We neglect the effects of refraction from density variations within the cold torus, or from the interface where the radiation exits the torus.  The refractive index outside the cold torus is assumed to be unity.  Inside the torus, the refractive index $n_{ct}$ depends on the orientation of the magnetic field relative to the radiation~\citep{gurnett}, but if we assume them to be parallel it is:

\begin{equation}
 n_{ct} = \sqrt{ 1 - \frac{\nu_p^2} {\nu (\nu - \nu_{cyc}) } }
 \label{eqn:refind}
\end{equation}

For a plasma density in the cold torus of $10^9$ cm$^{-3}$, compatible with \citet{Leto2006}, the plasma frequency is $\nu_p = 280$ MHz.  If we assume that the magnetic field strength at the point of refraction is approximately the same as at the point of emission, we have the cyclotron frequency $\nu_{cyc} = \nu / 2$ (if the emission is at the second harmonic).

\begin{figure*}
 \centering
 \begin{minipage}{0.45\linewidth}
  \centering
  \includegraphics[width=\linewidth]{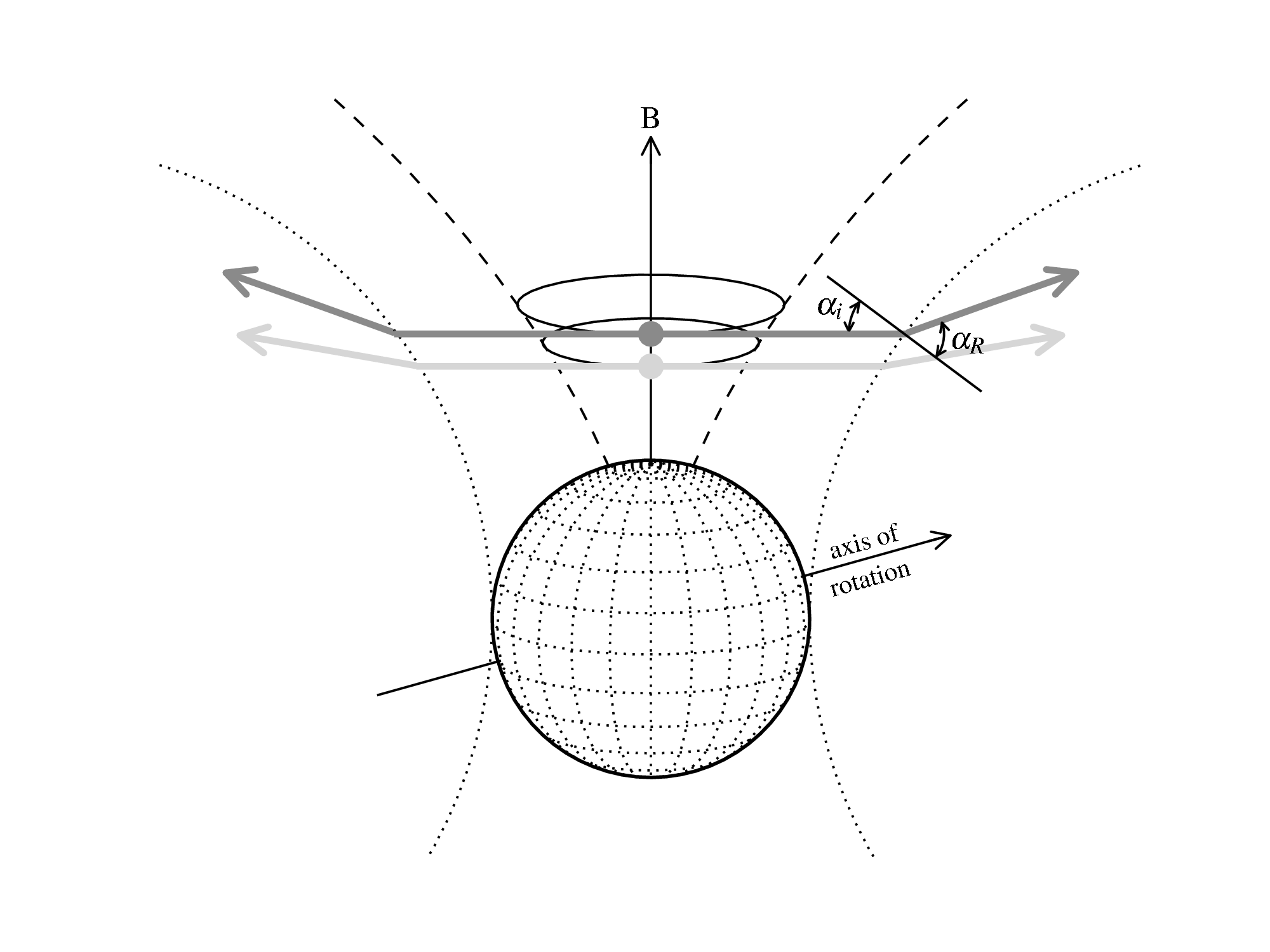}
 \end{minipage}
 \begin{minipage}{0.45\linewidth}
  \centering
  \includegraphics[width=\linewidth]{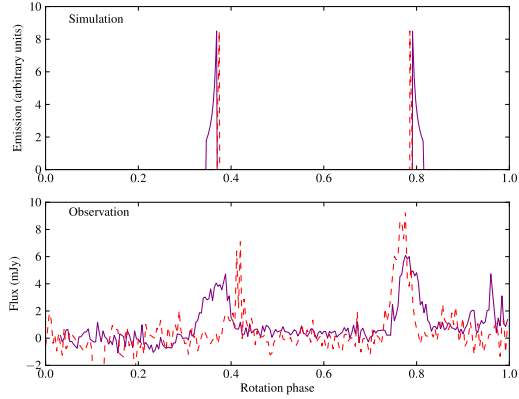}
 \end{minipage}
 \caption{Schematic diagram (left), and pulse profile of emission (right) for
the model of refracted ECM emission in the equatorial plane (section \ref{sec:tangentmodel}).
The regions in the magnetosphere where emission occurs are marked with solid
rings.  The emission in the 20\,cm and 13\,cm bands (dark and light arrows respectively, and solid and dotted lines on the right) is refracted as it crosses the boundary of the cold torus region (dotted).  This results in frequency-dependence in the pulse arrival time, corresponding with the observations (bottom right).  Note that the refraction in the left panel is exaggerated.}
 \label{fig:modelE}
\end{figure*}

If the angle of incidence $\alpha_i$ with the boundary of the torus (see Figure \ref{fig:modelE}) is assumed to be $60^\circ$, as by \citet{trigilio2011}, the angle of refraction $\alpha_R$ can be found with Snell's law:

\begin{equation}
 \sin \alpha_R = \frac{1}{n_{ct}} \sin \alpha_i
 \label{eqn:snell}
\end{equation}

For $n_{ct} < \sin \alpha_i$, this implies that the radiation is entirely reflected at the interface, and none is transmitted.  Under the conditions assumed here, this occurs for $\nu < 790$ MHz, so the pulse spectrum should be truncated below this frequency.  However, this is below the frequency range of our observations, so we cannot test this here.

Our simulation procedure is as follows:

\begin{enumerate}
 \item \label{it:tangent_start} Select a rotation phase and convert to magnetic latitude, as in section \ref{sec:paramodel}.

 \item Convert from the magnetic latitude of the emission to the angle of refraction.  From our assumption about the angle of incidence, this is done by adding $60^\circ$.

 \item Calculate the refractive index at this frequency with equation \ref{eqn:refind}.

 \item Determine the angle of incidence with equation \ref{eqn:snell}.

 \item Convert from the angle of incidence to the original magnetic latitude of the emission.  From our assumption about the angle of incidence, this is done by subtracting $60^\circ$.

 \item \label{it:tangent_getem} Determine the strength of emission in this direction.  We assume the emission to have a narrow Gaussian profile centred on the magnetic equator.

 \item \label{it:tangent_phase} Repeat steps \ref{it:tangent_start}-\ref{it:tangent_getem} for different rotation phases to form a pulse profile.

 \item \label{it:tangent_freq} Repeat steps \ref{it:tangent_start}-\ref{it:tangent_phase} for emission frequencies across the observation band.  Sum the resulting profiles, weighting them equally.
\end{enumerate}

The phase and height of the pulse profile are then adjusted as in section \ref{sec:paramodel}.  The Gaussian function used in step \ref{it:tangent_getem} is very narrow (width $< 0.2^\circ$), so the width of the resulting profile is due to variation with frequency across the observation band.

Results are shown in Figure \ref{fig:modelE} (right panel).  The pulse arrival times and their frequency dependence both fit the observations.  The observed pulses are wider than the simulated pulses; this may be due to inherent width of the emission beam, variation of the angle of incidence, or further turbulent refraction in the cold torus.

\section{Conclusions}

We observed CU Vir with the ATCA over six epochs and, in the 20\,cm
observing band, detected 100\% circularly polarised pulses twice in a
rotation period. This confirms that the 20\,cm pulses are stable over
a period of more than a decade.  We did not detect the 13\,cm leading
pulse that was observed a year earlier by \citet{rhw+10}. For our
data, the leading pulse has a spectrum which cuts off between 1.9 and
2.3\,GHz. We can see a clear frequency dependency in the time between
leading and trailing pulses across our entire observing band.

We have simulated in detail the emission models proposed by \citet{tll+00}, \citet{tlu+08} and \cite{trigilio2011}.  We show that the first two models do not fit our observations, the first being excluded by its lack of frequency dependence and the second by its grossly dissimilar pulse profile.  We confirm that the third model, involving refraction of the emission as it propagates through material in the stellar magnetosphere, yields the correct frequency dependence and pulse arrival times to fit our observations.

\section*{Acknowledgements}

The Australia Telescope Compact Array is part of the Australia
Telescope which is funded by the Commonwealth of Australia for
operation as a National Facility managed by CSIRO. K.L. is supported
by an Australian Postgraduate Award and a CSIRO OCE scholarship. K.L. and B.M.G. acknowledge the support of the Australian Research Council through grant DP0987072. The Centre for All-sky Astrophysics is an Australian Research Council Centre of Excellence, funded by grant CE11E0090. G.H. is the recipient of an
Australian Research Council QEII Fellowship
(no. DP0878388). V.R. is the recipient of a John Stocker Postgraduate Scholarship from the Science and Industry Endowment Fund. 
K.L. would like to thank Robert Braun for helpful
discussions. We are grateful to Bernie Walp for performing one of our observations.

\bibliography{journals,cuvir_refs}

\begin{thebibliography}{}

\bibitem[\protect\citeauthoryear{{Andre}, {Montmerle}, {Feigelson}, {Stine} \&
  {Klein}}{{Andre} et~al.}{1988}]{andre1988}
{Andre} P.,  {Montmerle} T.,  {Feigelson} E.~D.,  {Stine} P.~C.,    {Klein} K.,
   1988, ApJ, 335, 940

\bibitem[\protect\citeauthoryear{{Borra} \& {Landstreet}}{{Borra} \&
  {Landstreet}}{1980}]{borra+80}
{Borra} E.~F.,  {Landstreet} J.~D.,  1980, ApJS, 42, 421

\bibitem[\protect\citeauthoryear{{Deutsch}}{{Deutsch}}{1952}]{deutsch1952}
{Deutsch} A.~J.,  1952, ApJ, 116, 536

\bibitem[\protect\citeauthoryear{{Drake}, {Abbott}, {Bastian}, {Bieging},
  {Churchwell}, {Dulk} \& {Linsky}}{{Drake} et~al.}{1987}]{dab+87}
{Drake} S.~A.,  {Abbott} D.~C.,  {Bastian} T.~S.,  {Bieging} J.~H.,
  {Churchwell} E.,  {Dulk} G.,    {Linsky} J.~L.,  1987, ApJ, 322, 902

\bibitem[\protect\citeauthoryear{Gurnett \& Bhattacharjee}{Gurnett \&
  Bhattacharjee}{2005}]{gurnett}
Gurnett D.,  Bhattacharjee A.,  2005, Introduction to plasma physics.
Cambridge University Press

\bibitem[\protect\citeauthoryear{{Hatzes}}{{Hatzes}}{1997}]{hatzes1997}
{Hatzes} A.~P.,  1997, MNRAS, 288, 153

\bibitem[\protect\citeauthoryear{{Leone}}{{Leone}}{1991}]{Leone1991}
{Leone} F.,  1991, A\&A, 252, 198

\bibitem[\protect\citeauthoryear{{Leone}, {Trigilio} \& {Umana}}{{Leone}
  et~al.}{1994}]{Leone1994}
{Leone} F.,  {Trigilio} C.,    {Umana} G.,  1994, A\&A, 283, 908

\bibitem[\protect\citeauthoryear{{Leone} \& {Umana}}{{Leone} \&
  {Umana}}{1993}]{Leone1993}
{Leone} F.,  {Umana} G.,  1993, A\&A, 268, 667

\bibitem[\protect\citeauthoryear{{Leone}, {Umana} \& {Trigilio}}{{Leone}
  et~al.}{1996}]{lut+96}
{Leone} F.,  {Umana} G.,    {Trigilio} C.,  1996, A\&A, 310, 271

\bibitem[\protect\citeauthoryear{{Leto}, {Trigilio}, {Buemi}, {Umana} \&
  {Leone}}{{Leto} et~al.}{2006}]{Leto2006}
{Leto} P.,  {Trigilio} C.,  {Buemi} C.~S.,  {Umana} G.,    {Leone} F.,  2006,
  A\&A, 458, 831

\bibitem[\protect\citeauthoryear{{Linsky}, {Drake} \& {Bastian}}{{Linsky}
  et~al.}{1992}]{Linsky1992}
{Linsky} J.~L.,  {Drake} S.~A.,    {Bastian} T.~S.,  1992, ApJ, 393, 341

\bibitem[\protect\citeauthoryear{{Melrose} \& {Dulk}}{{Melrose} \&
  {Dulk}}{1982}]{melrose1982}
{Melrose} D.~B.,  {Dulk} G.~A.,  1982, ApJ, 259, 844

\bibitem[\protect\citeauthoryear{{North}}{{North}}{1998}]{north1998}
{North} P.,  1998, A\&A, 334, 181

\bibitem[\protect\citeauthoryear{{Pyper}, {Ryabchikova}, {Malanushenko},
  {Kuschnig}, {Plachinda} \& {Savanov}}{{Pyper} et~al.}{1998}]{prm+98}
{Pyper} D.~M.,  {Ryabchikova} T.,  {Malanushenko} V.,  {Kuschnig} R.,
  {Plachinda} S.,    {Savanov} I.,  1998, A\&A, 339, 822

\bibitem[\protect\citeauthoryear{{Ravi}, {Hobbs}, {Wickramasinghe}, {Champion},
  {Keith}, {Manchester}, {Norris}, {Bray} et~al.,}{{Ravi}
  et~al.}{2010}]{rhw+10}
{Ravi} V.,  {Hobbs} G.,  {Wickramasinghe} D.,  {Champion} D.~J.,  {Keith} M.,
  {Manchester} R.~N.,  {Norris} R.~P.,  {Bray} J.~D.,    et~al., 2010, MNRAS,
  408, L99

\bibitem[\protect\citeauthoryear{{Sault}, {Teuben} \& {Wright}}{{Sault}
  et~al.}{1995}]{sault+95}
{Sault} R.~J.,  {Teuben} P.~J.,    {Wright} M.~C.~H.,  1995, in {R.~A.~Shaw,
  H.~E.~Payne, \& J.~J.~E.~Hayes} ed., Astronomical Data Analysis Software and
  Systems IV Vol.~77 of Astronomical Society of the Pacific Conference Series,
  {A Retrospective View of MIRIAD}.
p.~433

\bibitem[\protect\citeauthoryear{{Trigilio}, {Leto}, {Leone}, {Umana} \&
  {Buemi}}{{Trigilio} et~al.}{2000}]{tll+00}
{Trigilio} C.,  {Leto} P.,  {Leone} F.,  {Umana} G.,    {Buemi} C.,  2000,
  A\&A, 362, 281

\bibitem[\protect\citeauthoryear{{Trigilio}, Leto, Umana, Buemi \&
  Leone}{{Trigilio} et~al.}{2008}]{tlu+08}
{Trigilio} C.,  Leto P.,  Umana G.,  Buemi C.~S.,    Leone F.,  2008, MNRAS,
  384, 1437

\bibitem[\protect\citeauthoryear{{Trigilio}, {Leto}, {Umana}, {Leone} \&
  {Buemi}}{{Trigilio} et~al.}{2004}]{Trigilio2004}
{Trigilio} C.,  {Leto} P.,  {Umana} G.,  {Leone} F.,    {Buemi} C.~S.,  2004,
  A\&A, 418, 593

\bibitem[\protect\citeauthoryear{{Trigilio}, {Leto}, {Umana}, {Simona Buemi} \&
  {Leone}}{{Trigilio} et~al.}{2011}]{trigilio2011}
{Trigilio} C.,  {Leto} P.,  {Umana} G.,  {Simona Buemi} C.,    {Leone} F.,
  2011, ArXiv e-prints: 1104.3268, Submitted to ApJ Letters

\bibitem[\protect\citeauthoryear{{Wilson}, {Ferris}, {Axtens}, {Brown},
  {Davis}, {Hampson}, {Leach}, {Roberts} et~al.,}{{Wilson}
  et~al.}{2011}]{wilson2011}
{Wilson} W.~E.,  {Ferris} R.~H.,  {Axtens} P.,  {Brown} A.,  {Davis} E.,
  {Hampson} G.,  {Leach} M.,  {Roberts} P.,    et~al., 2011, MNRAS, 416, 832

\end{thebibliography}
\bibliographystyle{mn2e}

\appendix

\section{Angular relations}
\label{sec:appendix}

\subsection{Rotation of \cuvir}
\label{sec:rot_appendix}

\begin{figure}
  \includegraphics[width=0.95\linewidth]{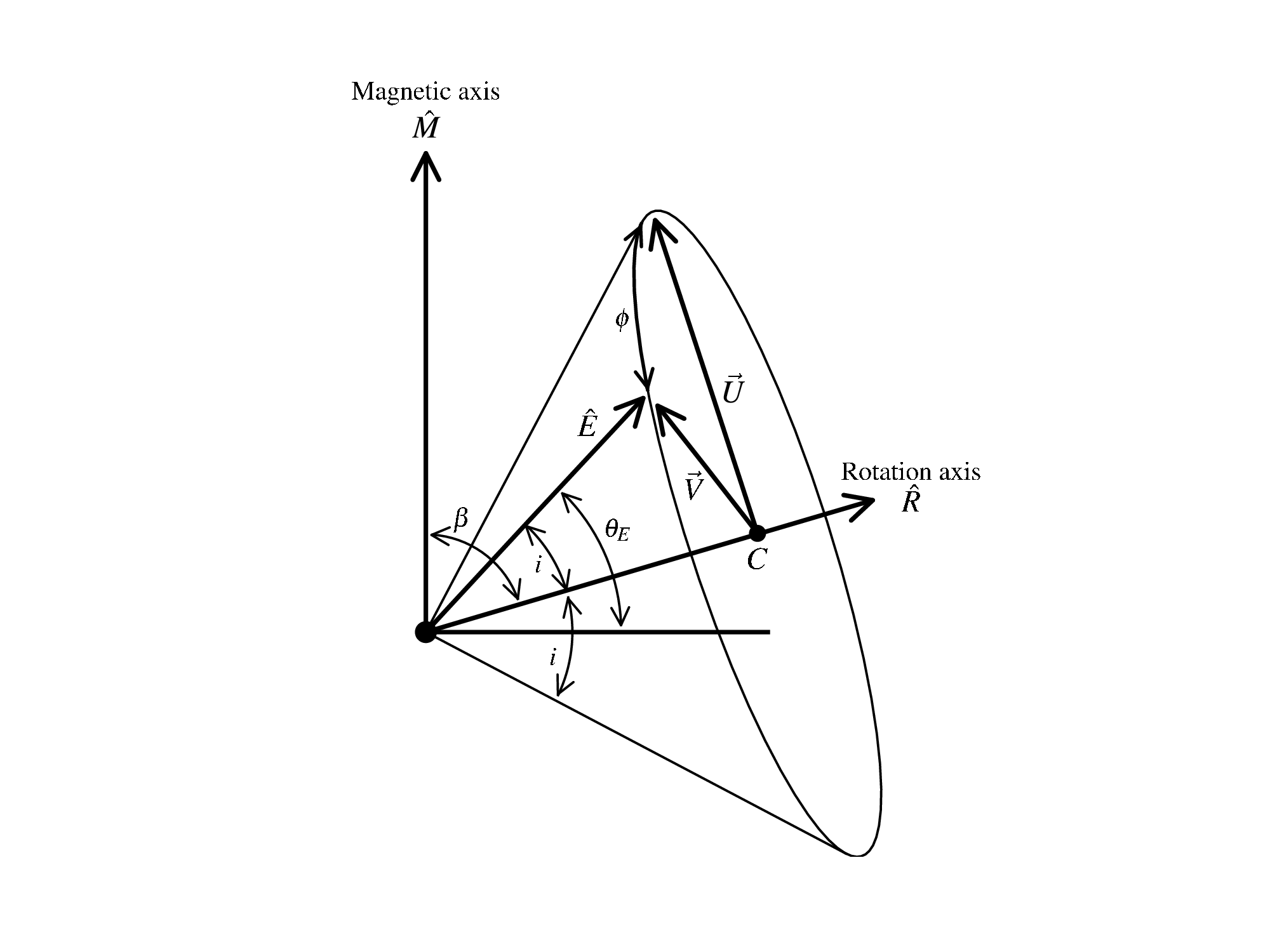}
  \caption{Rotation of \cuvir.  The magnetic axis $\hat{M}$ and the rotation axis $\hat{R}$ are inclined by $\beta$.  In the reference frame of the star, the vector towards the Earth $\hat{E}$ describes a circle around point $C$ on the rotation axis, separated from the rotation axis by $i$, and parameterised by the rotation phase $\phi$.  The magnetic latitude of this vector is $\theta_E$.  The vectors $\vec{U}$ and $\vec{V}$ are in the planes $\hat{R}$-$\hat{M}$ and $\hat{R}$-$\hat{E}$ respectively; for their application, see section \ref{sec:rot_appendix}.}
  \label{fig:rot_diagram}
\end{figure}

Equation \ref{eqn:rotation} connects the rotation phase of CU Vir to the magnetic latitude of emission in the direction of the Earth.  The relation between these is shown, with associated angles and unit vectors, in Figure \ref{fig:rot_diagram}.  To derive the relationship, we construct vectors:

\begin{align*}
\vec{U} &= \hat{E} - (\hat{E} \cdot \hat{S}) \hat{S} \\
\vec{V} &= \hat{B} - (\hat{B} \cdot \hat{S}) \hat{S}
\end{align*}

with lengths:

\begin{align*}
|\vec{U}| &= \sin i \\
|\vec{V}| &= \sin \beta
\end{align*}

The angle between these vectors is $\phi$, which we can then find as:

\begin{align}
\cos \phi &= \frac{\vec{U} \cdot \vec{V}} {|\vec{U}| |\vec{V}|} \nonumber \\
 &= \frac{\hat{E} \cdot \hat{B} - (\hat{E} \cdot \hat{S}) (\hat{S} \cdot \hat{B})} {\sin i \sin \beta}
\end{align}

As $\hat{E}$, $\hat{S}$ and $\hat{B}$ are unit vectors, we can evaluate these dot products as trigonometric identities:

\begin{equation}
 \cos \phi = \frac{\sin \theta_E - \cos i \cos \beta} {\sin i \sin \beta}
 \label{eqn:rot_reverse}
\end{equation}

Rearranging, we obtain:

\begin{equation}
 \sin\theta_E = \sin \beta \sin i \cos \phi + \cos \beta \cos i
\end{equation}

which is the same as equation \ref{eqn:rotation}.

\subsection{Conical emission}
\label{sec:cone_appendix}

Step \ref{it:dipole_decone} of the procedure in section \ref{sec:dipolemodel} assumes a relation, equation \ref{eqn:cone}, between the magnetic latitude of emission $\theta_E$ and the position $\omega$ of that emission on a cone, as shown in Figure \ref{fig:cone_diagram}.  To demonstrate this, we note that Figure \ref{fig:rot_diagram} is equivalent to Figure \ref{fig:cone_diagram}, with the following substitutions:

\begin{align*}
\beta & \rightarrow \frac{\pi}{2} - \theta_B \\
i & \rightarrow \delta \\
\phi & \rightarrow \omega
\end{align*}

Therefore the derivation of equation \ref{eqn:rot_reverse} in section \ref{sec:rot_appendix} also demonstrates the correctness of equation \ref{eqn:cone}:

\begin{equation}
\cos \omega =
 \frac{\sin \theta_E - \cos \delta \sin \theta_B}
      {\sin \delta \cos \theta_B}
\label{eqn:cone_rep}
\end{equation}

In step \ref{it:dipole_getem}, we then wish to find the emission power $P$ per solid angle $\Omega$ in this direction.  We may assume that the emission is evenly distributed around the cone (i.e. $dP/d\omega$ is constant).  If we differentiate equation \ref{eqn:cone_rep} with respect to $\theta_E$, we get:

\begin{equation}
\frac{d\omega}{d\theta_E} = - \frac{\cos\theta_E}{\cos\theta_B \sin\delta \sin\omega}
\end{equation}

This allows us to obtain the distribution of emission power with magnetic latitude:

\begin{align}
\frac{dP}{d\theta_E} &= \frac{dP}{d\omega} \frac{d\omega}{d\theta_E} \nonumber \\
 &\propto - \frac{\cos\theta_E}{\cos\theta_B \sin\delta \sin\omega} && \text{(as $\frac{dP}{d\omega}$ is constant)}
\end{align}

The emission power per solid angle is related to this:

\begin{align}
\frac{dP}{d\Omega} &= \frac{1}{2\pi} \frac{1}{\cos \theta_E} \frac{dP}{d\theta_E} \nonumber \\
 &\propto \frac{1}{\cos\theta_B \sin\delta \sin\omega}
\end{align}

This is the result used in step \ref{it:dipole_getem} of section \ref{sec:dipolemodel}: equation \ref{eqn:emission}.

\label{lastpage}
\end{document}